\documentclass[sigconf]{acmart}
\renewcommand\footnotetextcopyrightpermission[1]{}

\acmConference[ICSE 2024]{46th International Conference on Software Engineering}{April 2024}{Lisbon, Portugal}

\AtBeginDocument{%
  \providecommand\BibTeX{{%
    \normalfont B\kern-0.5em{\scshape i\kern-0.25em b}\kern-0.8em\TeX}}}

\setcopyright{acmcopyright}
\copyrightyear{2018}
\acmYear{2018}
\acmDOI{XXXXXXX.XXXXXXX}

\acmBooktitle{Woodstock '18: ACM Symposium on Neural Gaze Detection,
 June 03--05, 2018, Woodstock, NY} 
\acmPrice{15.00}
\acmISBN{978-1-4503-XXXX-X/18/06}

\usepackage{mdframed}
\usepackage{xcolor}
\usepackage{colortbl}
\usepackage{listings}
\usepackage{xspace}
\usepackage{multirow}
\usepackage{tikz}
\usepackage{pgfplots}
\usepackage{pgffor}
\usepackage{subcaption}
\usepackage{cleveref}
\usepackage{enumitem}
\usepackage{threeparttable}

\makeatletter
\DeclareRobustCommand\onedot{\futurelet\@let@token\@onedot}
\def\@onedot{\ifx\@let@token.\else.\null\fi\xspace}

\def\eg{\emph{e.g}\onedot}
\def\ie{\emph{i.e}\onedot}

\def\etc{\emph{etc}\onedot}

\makeatother

\definecolor{codegreen}{rgb}{0,0.6,0}
\definecolor{codegray}{rgb}{0.5,0.5,0.5}
\definecolor{codepurple}{rgb}{0.58,0,0.82}

\newcommand{\repeatcmd}[2]{\foreach \n in {1,...,#1}{#2}}
\newcommand{\spaces}[1]{\repeatcmd{#1}{\ }}
\newcommand{\gitadd}[2][1]{\textcolor{codegreen}{+\spaces{#1}#2}}
\newcommand{\gitdelete}[2][1]{\textcolor{red}{-\spaces{#1}#2}}
\newcommand{\code}[1]{{\ttfamily #1}}

\lstdefinestyle{javastyle}{
    basicstyle=\ttfamily\footnotesize,
    commentstyle=\color{codegreen},
    keywordstyle=\color{blue},
    numberstyle=\tiny\color{codegray},
    stringstyle=\color{codegreen},
    breakatwhitespace=false,         
    breaklines=true,                 
    captionpos=b,                    
    keepspaces=true,                 
    numbers=left,                    
    numbersep=1pt,                  
    showspaces=false,                
    showstringspaces=false,
    showtabs=false,                  
    tabsize=2
}

\newcommand{\cmt}{<Comment>\xspace}
\newcommand{\buggycode}{<Buggy Code>\xspace}
\newcommand{\fixedcode}{<Fixed Code>\xspace}
\newcommand{\bugloc}{[BUG\_LOCATION]\xspace}
\newcommand{\fixlocs}{[FIX\_START]\xspace}
\newcommand{\fixloce}{[FIX\_END]\xspace}
\newcommand{\cmtline}[1]{\rm\textcolor{codegray}{//\textbf{\cmt} #1}}

\newcommand{\gptt}{Chat\-GPT-3.5\xspace}
\newcommand{\gptf}{Chat\-GPT-4\xspace}
\newcommand{\llama}{LLaMA\xspace}
\newcommand{\codellama}{Code\-LLaMA\xspace}
\newcommand{\incoder}{InCoder\xspace}
\newcommand{\codetfp}{Code\-T5+\xspace}
\newcommand{\codegent}{Code\-Gen-2\xspace}
\newcommand{\codefuse}{Code\-Fuse\xspace}
\newcommand{\codereviewer}{Code\-Reviewer\xspace}

\begin{document}

\title{The Right Prompts for the Job: Repair Code-Review Defects with Large Language Model}

\author{Zelin Zhao}
\email{zelin.zzl@antgroup.com}
\affiliation{%
  \institution{Ant Group}
  \city{Hangzhou}
  \state{Zhejiang}
  \country{China}
}

\author{Zhaogui Xu}
\email{zhengrong.xzg@antgroup.com}
\affiliation{%
  \institution{Ant Group}
  \city{Hangzhou}
  \state{Zhejiang}
  \country{China}
}

\author{Jialong Zhu}
\email{zhujialong.zjl@antgroup.com}
\affiliation{%
  \institution{Ant Group}
  \city{Hangzhou}
  \state{Zhejiang}
  \country{China}
}
\affiliation{%
  \institution{Nanjing University}
  \city{Nanjing}
  \state{Jiangsu}
  \country{China}
}

\author{Peng Di}
\email{dipeng.dp@antgroup.com}
\affiliation{%
  \institution{Ant Group}
  \city{Hangzhou}
  \state{Zhejiang}
  \country{China}
}

\author{Yuan Yao}
\email{y.yao@nju.edu.cn}
\affiliation{%
  \institution{Nanjing University}
  \city{Nanjing}
  \state{Jiangsu}
  \country{China}
}

\author{Xiaoxing Ma}
\email{xxm@nju.edu.cn}
\affiliation{%
  \institution{Nanjing University}
  \city{Nanjing}
  \state{Jiangsu}
  \country{China}
}

\begin{abstract}
    Automatic program repair (APR) techniques have the potential to reduce manual efforts in uncovering and repairing program defects during the code review (CR) process. However, the limited accuracy and considerable time costs associated with existing APR approaches hinder their adoption in industrial practice. One key factor is the under-utilization of review comments, which provide valuable insights into defects and potential fixes. Recent advancements in Large Language Models (LLMs) have enhanced their ability to comprehend natural and programming languages, enabling them to generate patches based on review comments. This paper conducts a comprehensive investigation into the effective utilization of LLMs for repairing CR defects. In this study, various prompts are designed and compared across mainstream LLMs using two distinct datasets from human reviewers and automated checkers. Experimental results demonstrate a remarkable repair rate of 72.97\% with the best prompt, highlighting a substantial improvement in the effectiveness and practicality of automatic repair techniques.
\end{abstract}


\maketitle

\section{Introduction}\label{sec:intro}

In contemporary industrial practice, Continuous Integration/Continuous Deployment (CI/CD) pipelines are crucial in controlling the software development process. Within this pipeline, Code Review (CR) serves as a pivotal node where reviewers identify defects and provide comments, enabling developers to rectify the identified issues. 
Unfortunately, the CR process often demands significant efforts from reviewers and developers.
Automated checkers~\cite{pmd,infer,spotbugs} are widely employed to identify defects. The collaboration of human reviewers and automated checkers showcases the capability to identify a wide range of defects that exhibit varying complexities. 

The process of defect repair still heavily relies on manual efforts in practice. Automated Program Repair (APR) offers a fully automated solution that generally operates within the widely recognized GAV (Generate and Validate) paradigm \cite{apr-survey-2023-huang}.
Despite being fully automated, its inherent time-consuming nature makes integration within time-sensitive CI/CD pipelines challenging. Additionally, not all defects identified during code review manifest observable anomalous behaviors or result in failures. Hence, we propose a semi-automated paradigm of APR that leverages review comments to automatically generate patches, which are then presented to developers for manual validation.

Conventional approaches include \emph{search-based}, \emph{constraint-based}, and \emph{template-based methods} \cite{apr-survey-2023-huang}, which have shown practical applicability. However, these approaches encounter limitations when dealing with complex cases, potentially leading to incorrect patches or significant time overhead. Moreover, these approaches face challenges in effectively harnessing the valuable insights offered by review comments expressed in natural language. The inherent rigidity of these methodologies restricts their applicability and adaptability to a broader spectrum of defects. \emph{AI-based} \cite{apr-survey-2023-learning} APR approaches, specifically with the advancement of Large Language Models (LLMs), hold significant potential in comprehending both natural and programming languages, offering a promising solution to effectively address the underlying problem.

In this paper, we investigate the application of LLMs in repairing defects identified by reviewers and automated checkers during the CR process. Our study reveals that, for most CR defects, the associated review comments play a pivotal role in guiding the repair process. These comments not only explain the defect but often offer suggestions on how to rectify it. Similar to developers, pre-trained LLMs possess the capability to comprehend both natural and programming languages. Consequently, they can effectively analyze the review comments in conjunction with the faulty code and generate the appropriate and corrected version.

The effectiveness of a large language model in downstream tasks depends on the model base, prompts used, and the utilization of finetuning datasets. In order to explore effective methods for repairing CR defects, four research questions have been identified. Firstly, the study aims to determine the extent to which various LLMs can repair defects using zero-shot learning or finetuning. Secondly, the impact of different prompts containing varied information on LLM performance for CR defect repair is examined. Thirdly, the study investigates the performance of LLMs of different sizes to strike a balance between effectiveness and efficiency. Lastly, the research explores the potential of combining data from reviewers and automated checkers to enhance the repair capability of LLMs.

To answer the research questions, we studied nine different LLMs, including \gptt, \gptf, \llama, \codellama, and several other mainstream models. Seven different prompts were designed, encompassing various information related to the defects. Among these prompts, two were chosen for zero-shot learning, while five were utilized for finetuning purposes. In order to assess the significance of dataset diversity, two datasets of comparable size were collected, consisting of $\sim$16K reviewer comments and $\sim$15K automatically generated comments.

Our contributions are summarized as follows:
\begin{itemize}
    \item We proposed a semi-automated paradigm for APR in the CR process that harnesses the capabilities of LLMs to understand review comments and generate repair patches.
    \item We thoroughly studied the performance of 9 popular LLMs and observed that \codellama outperformed the others with a repair success rate of up to 72.97\%.
    \item We designed 7 prompts based on the information during the CR process and revealed that review comments and fix ranges were the two most effective prompts. The cross-validation of 2 distinct datasets demonstrated the importance of data diversity.
\end{itemize}

The rest is organized as follows. Section 2 overviews the background of code review. Section 3 describes the research questions and experiment settings. The experimental results are illustrated in section 4, which followed by the discussion of threats in section 5, related works in section 6 and the conclusion in section 7.

\section{Code Review}\label{sec:background}
\begin{figure*}[htb]
\centering
    \includegraphics[clip,width=1.0\textwidth]{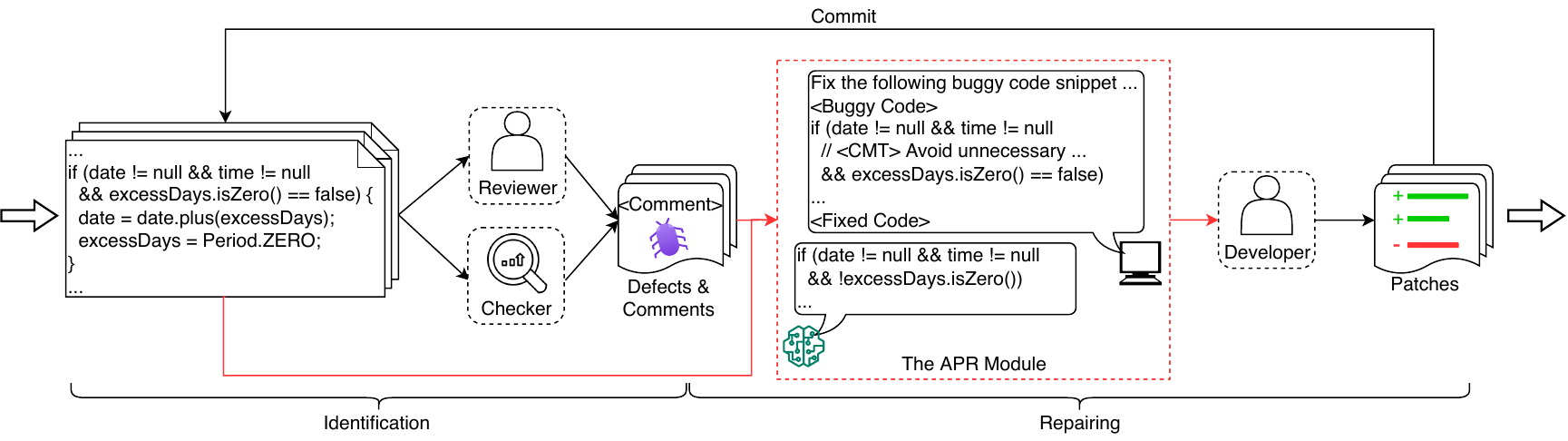}%
\caption{The code review process in industrial practice. Red lines denote the new steps and components to cooperate with CLM.}\label{fig:overview}
\vskip -2mm
\end{figure*}

Code Review plays a pivotal role in ensuring the quality of code changes integrated into the codebase. 
Typically, the CR process encompasses two fundamental steps: defect identification and repair. 
\Cref{fig:overview} provides an overview of the typical CR process practiced in industrial settings.

\subsection{Defect Identification}

The defect identification process, currently involving human reviewers and automated checkers, demonstrates the capability to identify various defects with varying complexities. Both reviewers and checkers contribute by providing comments describing the identified defects and, in some cases, offering suggestions on rectifying them. These crucial comments are vital in enhancing the effectiveness of following repair and optimization operations.

\begin{figure}[htb]
\small
\begin{subfigure}{\linewidth}
\begin{lstlisting}[language=Java,escapechar=|]
  public int getType() {
    if (code.startsWith("...warning")) {
      return TYPE_WARNING;
    } else if (code.startsWith("...disallowed")) {
      return TYPE_ERROR;
    } else if (info.startsWith("...AbuseFilter")) {
      return TYPE_ERROR;
    } else {
      |\cmtline{Why not simply return \code{TYPE\_ERROR} for this ``\code{else}'' case, and not worry about the contents of ``\code{info}''? This would prevent even more possible crashes.} \label{fig:code:1.a.9}|
|\gitdelete[6]{throw new RuntimeException("Unknown abusefilter response!");} \label{fig:code:1.a.10}|
|\gitadd[6]{return TYPE\_ERROR;} \label{fig:code:1.a.11}|
    }
 }
\end{lstlisting}
\vskip -2mm
\caption{The patch fixes the control flow and prevents more crashes. String literals in conditions are simplified.}\label{fig:code:1.a}
\end{subfigure}
\begin{subfigure}{\linewidth}
\begin{lstlisting}[language=Java,escapechar=|]
  public ListCardItemView setSubtitle(@Nullable CharSequence subtitle) {
    |\cmtline{Wouldn't this mean that we're calling the base method of \code{Text\\-View.setText()} and bypassing the overridden method?} \label{fig:code:1.b.2}|
|\gitdelete[4]{subtitleView.setText(subtitle);} \label{fig:code:1.b.3}|
|\gitadd[4]{subtitleView.setText((String) subtitle);} \label{fig:code:1.b.4}|
    return this;
  }
\end{lstlisting}
\vskip -2mm
\caption{The patch fixes a bug of incorrect method invocation.}\label{fig:code:1.b}
\end{subfigure}
\vskip -2mm
\caption{CR defects identified by reviewers.}\label{fig:code:1}
\end{figure}

Expert reviewers can identify intricate software semantic bugs. 
\Cref{fig:code:1} shows two observed defects in the CR process.
The accompanying comments are highlighted by the ``\cmt''  lines.
In \cref{fig:code:1.a}, 
the first reviewer discovered that throwing a \code{Run\-time\-Exception} results in more crashes and thus recommended addressing the defect by directly returning \code{TYPE\_ERROR}.
In \cref{fig:code:1.b}, the second reviewer revealed that instead of the overridden one, the base method would be invoked because the code in \cref{fig:code:1.b.3} does not cast type.
Notably, the reviewer merely expressed a cautious concern about this anomalous behavior.

Automated checkers, such as PMD~\cite{pmd}, Infer~\cite{infer}, and Spotbugs~\cite{spotbugs}, offer the advantage of swiftly detecting common defects based on predefined patterns.
These tools are also capable of generating comments that describe the identified defects.
\Cref{fig:code:2} presents two defects identified by the PMD tool.
In \cref{fig:code:2.a}, the \code{Date\-Formatter} type is not thread-safe when accessed by multiple threads.
To address this concurrent defect, the corresponding patch introduces a synchronized block.
In \cref{fig:code:2.b}, comparing the boolean value returned from the \code{is\-Zero} method with the \code{false} literal is unnecessary.
Consequently, the patch simplifies the code and improves its readability as well.

\subsection{Repairing}

\begin{figure}[htb]
\small
\begin{subfigure}{\linewidth}
\begin{lstlisting}[language=Java,escapechar=|]
  static Date parse(String day) throws ParseException {
    |\cmtline{Static \code{DateFormatter} objects should be accessed in a synchronized manner.} \label{fig:code:2.a.2}|
|\gitadd[4]{synchronized(DAY\_STORAGE\_FORMAT) \{} \label{fig:code:2.a.3}|
      return DAY_STORAGE_FORMAT.parse(day);
|\gitadd[4]{\}} \label{fig:code:2.a.5}|
  }
\end{lstlisting}
\vskip -2mm
\caption{The patch fixes a possible concurrency defect.}\label{fig:code:2.a}
\end{subfigure}
\begin{subfigure}{\linewidth}
\begin{lstlisting}[language=Java,escapechar=|]
  private void resolvePeriod() {
    if (date != null && time != null
       |\cmtline{Avoid unnecessary comparisons in boolean expressions.} \label{fig:code:2.b.3}|
|\gitdelete[8]{\&\& excessDays.isZero() == false) \{} \label{fig:code:2.b.4}|
|\gitadd[8]{\&\& !excessDays.isZero()) \{} \label{fig:code:2.b.5}|
      date = date.plus(excessDays);
      excessDays = Period.ZERO;
    }
  }
\end{lstlisting}
\vskip -2mm
\caption{The patch simplifies the code and improves the readability.}\label{fig:code:2.b}
\end{subfigure}
\vskip -2mm
\caption{CR defects identified by automated checker, PMD.}\label{fig:code:2}
\end{figure}


In practice, the CR comments play a crucial role in assisting developers in pinpointing the specific locations of defects and devising suitable patches.
Once developers have implemented the necessary fixes, they proceed to commit the patch to the codebase,
triggering reviewers and checkers to engage in another round of code review to verify the absence of any lingering defects.

Presently, defect repair predominantly relies on manual effort. 
The prevailing process of APR offers a fully automated solution, which commonly operates within the well-known GAV paradigm.
Given a set of tests that expose bugs, APR algorithms first identify potential bug locations,
then generate a multitude of patch candidates for each identified location,
and finally validate each generated patch through the execution of the tests mentioned above
until either a test-passing patch is discovered or a predefined time limit is reached.
While this paradigm is fully automated, their time-consuming nature renders them unsuitable for integration within practical, time-sensitive CI/CD pipelines.
Furthermore, not all defects identified during code review directly manifest observable anomalous behaviors or trigger failures.
For example, the defect in \cref{fig:code:2.b} violates coding style and can not cause any failures, while the defect in \cref{fig:code:2.a} is a concurrency bug requiring robust tests to expose stably.

This paper proposes a semi-automated paradigm to leverage APR techniques effectively in the CR process. As illustrated in Figure \ref{fig:overview}, an APR module in the pipeline takes as input the comprehensive information available in the CR process, including code snippets and associated comments from both the reviewers and checkers. It automatically generates patch suggestions tailored to the identified code issues. These patch suggestions and the original comments are then presented to developers for consideration and further action. By adopting a semi-automated solution, developers can benefit from the assistance provided by APR techniques in generating potential fixes for identified code issues. This collaborative approach combines developers' expertise with APR techniques' capabilities, leading to efficient and effective software development process.

Many repair methodologies have been proposed so far. 
Traditional approaches, such as \emph{search-based}, \emph{constraint-based}, and \emph{template-based} methods, can be adopted in practice. 
Nevertheless, these approaches are primarily effective in resolving simple defects, as evident in the case in \cref{fig:code:2.b}.
Regrettably, when confronted with complicated or less prevalent cases, they have limitations that may result in incorrect patches or high time costs. Furthermore, these approaches need help to effectively utilize the valuable information provided by CR comments in natural language. 
Their inherent rigidity hinders their applicability and adaptability to a broader range of defects, such as the case shown in \cref{fig:code:1.a} and \cref{fig:code:1.b}. 
\emph{AI-based} APR approaches exhibit significant promising prospects. Especially with the development of LLMs, their capability of comprehending both program languages and natural language offers a prospective solution that can effectively tackle the underlying problem.
CR comments provide valuable insight into the defect and even the critical hint of the correct patch in natural language. 
LLMs can effectively acquire this knowledge, and when combined with the codebase, they can propose suitable fixes. We have made a comprehensive experiment on the APR with various LLMs. Our experiment demonstrates that LLMs can successfully suggest accurate fixes for the issues depicted in \cref{fig:code:1.a} and \cref{fig:code:1.b}, utilizing the reviewer comments as their prompts.

\section{Research Questions and Experiment Settings}\label{sec:method}
This section describes the research questions and experiment settings on the APR with LLMs. The effectiveness of a large language model in performing a downstream task is typically influenced by three crucial factors: the underlying model base, the choice of prompts, and the utilization of finetuning datasets. Considering these factors, we have formulated the following research questions to guide our investigation of the automatic program repair task:

\begin{itemize}[leftmargin=*, noitemsep, topsep=0pt]
    \item \textbf{RQ1}: To what extent can various LLMs repair CR defects using zero-shot learning or finetuning?
    \item \textbf{RQ2}: How does the choice of prompts impact the performance of LLMs in repairing defects?
    \item \textbf{RQ3}: How does the performance of LLMs in repairing defects vary with different model sizes?
    \item \textbf{RQ4}: How do different datasets affect the capacity to rectify defects, and is it feasible to interchangeably employ these datasets?
\end{itemize}

In the following section, we outline the specific details of our experimental setup for addressing the above research questions.

\subsection{Models}

Large language models display variations in their architectures and parameter quantities and undergo pretraining with different datasets. As a result, their performances also differ when they are applied to different downstream tasks.

LLMs mainly follow the transformer architecture in three different ways. 
\emph{Encoder-only} models, like BERT \cite{bert}, rely solely on Transformer encoders and are typically trained to generate neural embeddings of data by predicting masked tokens from original inputs.
\emph{Decoder-only} models, like GPTs \cite{chatgpt}, exclusively consist of Transformer decoders and are trained by predicting the next token based on all preceding tokens. These models are suitable for generative tasks.
\emph{Encoder-Decoder} models, like T5 \cite{t5}, employ the full Transformer architecture and are usually trained by restoring the original sequence from corrupted inputs (\eg, span masking), which can be utilized in both tasks mentioned above.

There are two primary approaches for accomplishing specific tasks, namely \emph{zero-shot} learning and \emph{finetuning}. 
Zero-shot learning directly utilizes the underlying model for tasks, which typically necessitates a stronger inherent capability of the model. In contrast, finetuning relies on a collection of high-quality labeled datasets to additionally train the base model. In the context of the APR task, selecting an appropriate model base is crucial. To evaluate the performance, we meticulously handpicked a few models from the mainstream. These models were specifically chosen for their demonstrated proficiency in comprehending both natural language and programming languages, as well as their ability to generate high-quality code snippets. In the following, we list the LLMs we chose for our experiment.

\begin{itemize}[leftmargin=*, noitemsep, topsep=0pt]
    \item \textbf{ChatGPT \cite{chatgpt}:} ChatGPTs are widely recognized for their awe-inspiring generation capabilities in a conversational manner, resembling human-like interactions. ChatGPT 3.5 175B was released in November 2022, followed by ChatGPT 4 in March 2023. For our experiment, we selected both ChatGPT 3.5 and ChatGPT 4. We utilize them for zero-shot learning as the task baseline through their open APIs.
    \item \textbf{\incoder \cite{incoder-23}:} \incoder is a decoder-only Transformer model designed for code infilling and synthesis. This model is pretrained with a causal masking objective on 159 GB of open-source code from GitHub and GitLab and 57 GB of questions and answers from StackOverflow. The official release includes two variants, one with 1B and another with 6B parameters. We considered the latter one for our experiment.
    \item \textbf{\codetfp \cite{codet5p-23}:} \codetfp is a member of encoder-decoder Transformer models. It is specifically designed to enhance both code understanding and generative tasks. It has been trained on a large dataset of 51.5B tokens, encompassing both unimodal and bimodal data. The CodeT5+ family comprises five distinct model sizes, ranging from 220M to 16B parameters. For comparison with other models, we selected the 6B version. Furthermore, we also conducted additional experiments to investigate the impact of different model sizes on performance within this model family.
    \item \textbf{\codefuse \cite{codefuse}:} \codefuse, developed by AntGroup Inc., is a member of decoder-only Transformers, designed for producing accurate, efficient, and compliant code. It is pretrained on a 1000B token code dataset covering over 40 programming languages. The model size varies from 6B to 34B. We chose the 6B version for our experiment.
    \item \textbf{\llama \cite{llama-23}:} \llama, released by Meta Inc., is also a decoder-only Transformer model. The training dataset contains about 1.4T tokens extracted from various sources, including CommonCrawl, GitHub, Wikipedia, \etc.
    We considered the model with 7B parameters in our experiment.
    \item \textbf{\codegent \cite{codegen2-23}:} \codegent is a Prefix-LM-based Transformer model designed to unify common training schemes, such as model architecture and learning algorithm. It has been trained with a mixed objective of causal language modeling and span corruption on both natural language and source code. We utilized the 7B version in our experiments.
    \item \textbf{\codellama \cite{codellama-23}:} \codellama is also a decoder-only Transformer model derived from LLaMA-2. It has three variants: the base one is additionally trained on a code-heavy dataset based on LLaMA-2, and the other two are further finetuned on Python and instruction datasets, respectively. We selected the base variant with 7B parameters for our experiments. 
    \item \textbf{\codereviewer \cite{codereviewer-22}:} \codereviewer is an  encoder-decoder Transformer model explicitly designed for the code review domain. It is trained on annotated code changes sourced from GitHub repositories. The model consists of 220 million parameters, making it comparatively smaller. Notably, the model lacks training on natural language, so we refrain from finetuning this model with prompts and instead observe the results within its input-output format.
\end{itemize}

In our experiment, we utilized the library of HuggingFace transformers to deploy the models and incorporated techniques such as Low-Rank Adaptation and Mixed Precision Training. The finetuning process was performed using 2 NVIDIA A100-SXM4-80GB GPUs, 200 GB of memory, and an AMD EPYC 7T83 CPU. Consistency was maintained across all models by adopting the same hyperparameters, including two training epochs, a learning rate of 2e-4, and a batch size of 16.

\subsection{Prompts}

Prompts serve as the primary means by which users engage with LLMs, establishing a crucial interface for interaction. The design of prompts holds significant influence over the resultant inferences, rendering it a vital aspect to consider when employing LLMs for specific tasks. Identifying the optimal prompt that maximizes performance while minimizing implementation effort becomes imperative. To compare the effects of different available information for the APR task, we design a set of prompts following the template shown in \cref{fig:prompt:pattern}.
Each prompt contains a subset of information in this figure. We aim to systematically assess the effects of differing information subsets on LLM performance.

\begin{figure}[htb]
\small
\begin{lstlisting}[language=Java,escapechar=|]
  |\rm \textcolor{codegray}{\textbf{Fixing the following buggy code snippet ...}}|
  |\rm \textcolor{codegray}{\textbf{\buggycode}}|
  public static Integer get(String key) {
    |\cmtline{Can this key ever get null?}|
    |\rm \textcolor{codegray}{\textbf{\bugloc}}|
    |\rm \textcolor{codegray}{\textbf{\fixlocs}}|
    if (key == null) {
      return R.drawable.acc_none;
    }
    |\rm \textcolor{codegray}{\textbf{\fixloce}}|
    return ...;
  } 
  |\rm \textcolor{codegray}{\textbf{\fixedcode}}|
\end{lstlisting}
\vskip -2mm
\caption{The template of prompts in this paper.}\label{fig:prompt:pattern}
\end{figure}

In addition to including the buggy code snippet, all prompts consist of three fundamental elements. First, the task instruction is presented in the initial line, providing a concise description of the code repair task and pertinent details regarding the specific settings and any additional requirements relevant to the desired inference result.
Second, the \emph{``\buggycode''} marker is employed in the second line to indicate that the subsequent code snippet contains bugs or errors that necessitate correction.
Third, the \emph{``\fixedcode''} marker is employed in the final line of the prompt, signifying the expectation for the model to generate an inference of the fixed version of the buggy code snippet. 
Overall, we have prepared two categories of prompts, one for zero-shot learning and the other for finetuning purposes.

\subsubsection{Prompts for Zero-shot Learning}
As a baseline, we designed two prompts of zero-shot learning for \gptt and \gptf.
\begin{itemize}[leftmargin=*, noitemsep, topsep=0pt]
   \item \textbf{Prompt 1 (P1):} It only contains basic information. The task instruction is \emph{``Fix the following buggy code snippet. In your response, output the fixed code only.''}. Note that ChatGPTs often generate natural language content to explain the bug or the patch, so we add the extra requirement in the task instruction to reduce such explanation.
   \item \textbf{Prompt 2 (P2):} It additionally includes the review comment as the critical hint. The task instruction is \emph{``Fix the following buggy code snippet according to the suggestion in the "//\cmt" line. In your response, output the fixed code only.''}. We ask ChatGPTs to fix the bug according to the review comment after the \emph{``\cmt''} marker.
\end{itemize}

\subsubsection{Prompts for Finetuning}
We designed five prompts to finetune other selected LLMs.

\begin{itemize}[leftmargin=*, noitemsep, topsep=0pt]
\item \textbf{Prompt 3 (P3):} It is a variant of Prompt 1, which does not have any hint of the defect within it. The task instruction is \emph{``Fix the following buggy code snippet.''}. The extra requirement is removed because the finetuning process enables the model to learn the correct behavior, \ie, outputting code only.
\item \textbf{Prompt 4 (P4):} It additionally provides the defect location. It indicates the buggy code line from the reviewer comments or the checker inspections. This line is denoted with \emph{``\bugloc''}. The task instruction is \emph{``Fix the following buggy code snippet. \bugloc marks the bug location.''}.
\item \textbf{Prompt 5 (P5):} It provides the suggested fix location, \ie, the line range where the code should be changed to fix the defect. Note that the fix location is not always the same as the defect location. Human review comments commonly present a suggested fix at certain code lines. The markers \emph{``\fixlocs''} and \emph{``\fixloce''} are denoted as the start and end of the patch, respectively. The task instruction is \emph{``Fix the following buggy code snippet. \fixlocs and \fixloce mark the range of the patch.''}. 
\item \textbf{Prompt 6 (P6):} It is an advanced version of Prompt 4.
It provides the review comment to help the LLM understand and repair the defect. Compared to Prompt 4, the comment line indicates the defect location with no special markers. It is designed to evaluate whether the rich comments content can improve the performance compared to the mere defect locations. The task instruction is \emph{``Fix the following buggy code snippet according to the suggestion in the "//\cmt" line.''}. 
\item \textbf{Prompt 7 (P7):} It is the combination of all information in \cref{fig:prompt:pattern} except \emph{``\bugloc''}. We eliminate the defect location because it is often implicit in the review comment. Note that compared to all prompts, this one is the most informative. The task instruction is \emph{``Fix the following buggy code snippet according to the suggestion in the "//\cmt" line. \fixlocs and \fixloce mark the range of the patch.''}.
\end{itemize}

\subsection{Datasets}

We collected two kinds of high-quality datasets for finetuning. One is from real-world reviewer comments, and the other is from the PMD~\cite{pmd} checker. Each instance of the dataset consists of the following information.

\begin{itemize}
    \item the buggy code snippet;
    \item the fixed code snippet;
    \item the comment either from reviewers or checkers;
    \item the defect location;
    \item the code range for fixing the defect.
\end{itemize}

\subsubsection{Dataset from Reviewer Comments (RD)}
In our experiment, we used the dataset provided by Tufano \etc~\cite{hd-21}. This dataset consisted of approximately 17,000 reviewer comments gathered from GitHub and Gerrit. However, we performed a manual inspection to enhance the data's quality and applied specific pruning rules to eliminate irrelevant or low-quality comments. The following rules were followed during the data pruning process:
\begin{itemize}
    \item The code snippet should not be changed too significantly. For example, those even no intersection between the buggy and fixed code snippet were eliminated;
    \item The comment should be meaningful. For example, the meaningless comments ``done'' or ``fixed'' were eliminated.
\end{itemize}

After all, we obtained a set of 16,228 data entries for finetuning.

\subsubsection{Dataset from the PMD Checker (PD)}

We utilized the PMD checker to collect the other kind of dataset. PMD is a powerful static rule-based checker that automatically examines source code in various programming languages and provides valuable feedback to developers. It is open-sourced and widely used in modern practices. In our experiment, we selected 30 rules frequently identified in real-world coding practices.

To construct our dataset, we performed scans on a collection of popular open-source Java projects hosted on GitHub. From these projects, we carefully handpicked the target buggy code snippets. To maintain a similar magnitude to the dataset from reviewer comments, we ultimately chose 20,000 code snippets that violated the selected 30 rules. In order to provide the corresponding fixes for these code snippets, we engaged the expertise of six experienced developers. They utilized the comments generated by PMD to guide them in fixing the identified defects. The fixes were meticulously reviewed to ensure their accuracy and effectiveness. After this process, our dataset comprised a total of 14,935 data entries.

When collecting the two datasets, we did not utilize code and comments from AntGroup internally. This is mainly because of the confidentiality requirements. Additionally, the majority of reviewer comments in AntGroup are written in Chinese. The pretraining datasets of some models only include a small portion of Chinese. Finetune these models with English comments is more reasonable.

In our experiment, the two datasets is divided into training and validation data in a 9:1 proportion. More specifically, the training data consists of 28,063 samples, while the validation data comprises 3,100 samples.

\subsection{Result Validation}

Executing test cases is commonly used to validate patches generated by APR techniques. However, in our specific scenario, this approach is unsuitable due to the nature of the results generated by LLMs. These results consist of incomplete code snippets, and as mentioned before, some CR defects do not exhibit observable runtime behaviors, rendering validation through tests impossible. Directly comparing code snippets as strings can lead to inaccurate judgments. LLMs often introduce or remove whitespace in code snippets and may even add comments to explain the fix, mainly observed in ChatGPTs. These differences can significantly impact the evaluation results, making direct string comparisons impractical and unreliable.

In our experiment, we have adopted two primary metrics for evaluation purposes. Firstly, we propose the \emph{exact-code-match (ECM)} metric to compare the generated code snippet with the expected one. The implementation of this metric involves using TreeSitter \cite{treesitter} to parse a code snippet into a partial abstract syntax tree (AST), then removing non-code elements (such as comments) from the AST, and finally printing the AST as formatted code. By directly comparing the formatted code of the generated and expected snippets, we can determine if they match exactly. This approach enables a more precise and meaningful code comparison, resulting in more accurate evaluation results. Secondly, we calculate the Code BLEU~\cite{code-bleu} between the generated result and the expected snippet. It is important to note that when calculating the Code BLEU in our experiments (as described in \cref{sec:experiment}), we use the output result directly without removing the non-code elements. The ECM metric does not consider non-code elements and evaluates the genuine capability of generating fixed code for CR defects. On the other hand, the Code BLEU metric considers the raw output and evaluates the overall performance of each model.

\section{Experiment results}\label{sec:experiment}

In this section, we individually demonstrate our experiment results for the proposed research questions. Recall our research questions as the following:
\begin{itemize}[leftmargin=*, noitemsep, topsep=0pt]
    \item \textbf{RQ1}: To what extent can various LLMs repair CR defects using zero-shot learning or finetuning?
    \item \textbf{RQ2}: How does the choice of prompts impact the performance of LLMs in repairing defects?
    \item \textbf{RQ3}: How does the performance of LLMs in repairing defects vary with different model sizes?
    \item \textbf{RQ4}: How do different datasets affect the capacity to rectify defects, and is it feasible to interchangeably employ these datasets?
\end{itemize}

\subsection{Overall Effectiveness (RQ1)}

\subsubsection{Results of Zero-shot Learning}

\Cref{table:baseline} shows the results of fixing CR defects by \gptt and \gptf through zero-shot learning. When considering the most straightforward prompt (P1), it is evident that \gptt outperforms \gptf in terms of fixing defects. Specifically, the ECM for \gptt is 14.10\%, while the ECM for \gptf is only 6.61\%. Prompt P1 instructs both models to fix the buggy code snippet without accompanying reviewer comments. Consequently, both models rely exclusively on their pretraining knowledge to detect potential defects and generate the corresponding corrected code. This reliance on pretraining knowledge introduces a potential limitation, as it may undermine the meaningfulness of the results.

\begin{table}[htb]
    \caption{Results of ChatGPTs in zero-shot learning.}
    \label{table:baseline}
    \centering
    \small
\begin{tabular*}{\columnwidth}{@{\extracolsep{\fill}} c|cc|cc}
    \toprule
    \multirow{2}{*}{\textbf{Prompt}} & \multicolumn{2}{c|}{\textbf{ECM}} & \multicolumn{2}{c}{\textbf{Code BLEU}}  \\ 
    & \gptt & \gptf & \gptt & \gptf \\
    \midrule
    P1 & 14.10\% & 6.61\% & 0.7789 & 0.7012  \\
    P2 & 49.42\% & 55.10\% & 0.8554 & 0.8867  \\
    \bottomrule
\end{tabular*}
\end{table}

\begin{table*}[htb]
\begin{threeparttable}
    \caption{The Performance of different models in the finetuning setting.
    The last row ($\textbf{\#}_{\textbf{P7}}$-$\textbf{\#}_{\textbf{P3}}$) shows the performance improvement from P3 to P7.}
    \label{table:main}
    \centering
\begin{tabular}{@{}p{\textwidth}@{}}
    \centering
    \small
\begin{tabular}{wc{0.8cm}|wc{0.9cm}wc{0.9cm}wc{0.9cm}wc{0.9cm}wc{0.9cm}wc{1.3cm}|wc{0.9cm}wc{0.9cm}wc{0.9cm}wc{0.9cm}wc{0.9cm}wc{1.3cm}}
    \toprule
    \multirow{2}{*}{\textbf{Prompt}} & \multicolumn{6}{c|}{\textbf{ECM}} & \multicolumn{6}{c}{\textbf{Code BLEU}} \\
    & \incoder & \codetfp & \codefuse & \llama & \codegent & \codellama  & \incoder & \codetfp & \codefuse & \llama & \codegent & \codellama \\
    \midrule
    P3         & \cellcolor{gray!10}39.87\%	& \cellcolor{gray!20}41.39\%	& \cellcolor{gray!40}44.26\%	& \cellcolor{gray!30}42.84\%	&\cellcolor{gray!50}45.61\%	& \cellcolor{gray!60}47.87\% & \cellcolor{gray!20}0.8635	& \cellcolor{gray!10}0.8625	& \cellcolor{gray!30}0.8658	& \cellcolor{gray!50}0.8817	& \cellcolor{gray!40}0.8708	& \cellcolor{gray!60}0.8861	\\
    P4         & \cellcolor{gray!10}45.06\%	& \cellcolor{gray!20}46.65\%	& \cellcolor{gray!40}49.71\%	& \cellcolor{gray!30}49.42\%	&\cellcolor{gray!50}51.16\%	& \cellcolor{gray!60}54.48\% & \cellcolor{gray!10}0.8737	& \cellcolor{gray!20}0.8767	& \cellcolor{gray!30}0.8788	& \cellcolor{gray!50}0.8946	& \cellcolor{gray!40}0.8835	& \cellcolor{gray!60}0.9015	\\
    P5         & \cellcolor{gray!10}52.61\%	& \cellcolor{gray!20}53.68\%	& \cellcolor{gray!30}56.48\%	& \cellcolor{gray!40}56.90\%	&\cellcolor{gray!50}60.00\%	& \cellcolor{gray!60} 62.97\% & \cellcolor{gray!10}0.8867	& \cellcolor{gray!20}0.8873	& \cellcolor{gray!30}0.8911	& \cellcolor{gray!50}0.9100	& \cellcolor{gray!40}0.8987	& \cellcolor{gray!60}0.9222	\\
    P6         & \cellcolor{gray!20}57.81\%	& \cellcolor{gray!10}57.65\%	& \cellcolor{gray!30}60.13\%	& \cellcolor{gray!40}60.55\%	&\cellcolor{gray!50}61.48\%	& \cellcolor{gray!60}66.39\% & \cellcolor{gray!30}0.8961	& \cellcolor{gray!10}0.8912	& \cellcolor{gray!20}0.8949	& \cellcolor{gray!50}0.9141	& \cellcolor{gray!40}0.8996	& \cellcolor{gray!60}0.9245	\\
    P7         & \cellcolor{gray!10}62.52\%	& \cellcolor{gray!20}63.03\%	& \cellcolor{gray!30}65.77\%	& \cellcolor{gray!40}66.48\%	&\cellcolor{gray!50}67.87\%	& \cellcolor{gray!60}72.97\% & \cellcolor{gray!20}0.9041	& \cellcolor{gray!10}0.9011	& \cellcolor{gray!30}0.9050	& \cellcolor{gray!50}0.9237	& \cellcolor{gray!40}0.9115	& \cellcolor{gray!60}0.9394	\\
    \midrule
    $\#_{P7}$-$\#_{P3}$      & \cellcolor{gray!40}22.65\%	&\cellcolor{gray!20}21.64\% 	& \cellcolor{gray!10}21.51\%	& \cellcolor{gray!50}23.64\%	&\cellcolor{gray!30}22.26\%	& \cellcolor{gray!60}25.10\% & \cellcolor{gray!30}0.0406	& \cellcolor{gray!10}0.0386	& \cellcolor{gray!20}0.0392	& \cellcolor{gray!50}0.042	& \cellcolor{gray!40}0.0407	& \cellcolor{gray!60}0.0533 \\
    \bottomrule
\end{tabular}
\end{tabular}
\begin{tablenotes}[flushleft]\footnotesize
    \item In the same row, the darker the color, the higher the data. The ECM and Code BLEU of all models increase from P3 to P7. Because of the particular data format and smaller model size, we only finetuned the \codereviewer model with review comments, similar to P2 and P6. The ECM and Code BLEU of \codereviewer are 47.87\% and 0.8444.
\end{tablenotes}
\end{threeparttable}
\end{table*}

Regarding the prompt P2, which provides review comments to the models during the fixed code generation, \gptt and \gptf exhibit significant improvements compared to the P1 setting. Specifically, \gptt achieves an ECM of 49.42\%, while \gptf achieves an ECM of 55.10\%. Such improvements suggest that review comments are crucial in effectively repairing CR defects. Furthermore, it is observed that \gptf surpasses \gptt in terms of performance in this setting. This can be attributed to the fact that \gptf possesses a deeper understanding of review comments compared to \gptt. The Code BLEU metric presented in \cref{table:baseline} corroborates the effectiveness of the ECM.

The performances of \gptt and \gptf using prompt P2 demonstrate their effectiveness in repairing CR defects. However, it is worth noting that these models may occasionally generate unexpected content, even when instructed to output only the fixed code. For instance, both \gptt and \gptf may attempt to complete the code snippet instead of generating the necessary fixes. Additionally, they may generate natural language content that describes the identified defect or explains the fix. While this can be informative, it often requires human intervention to prune and refine the generated content to ensure its accuracy and relevance.

\subsubsection{Results of Finetuning}

\Cref{table:main} presents each model's ECM and Code BLEU scores, excluding \codereviewer, after finetuning with various prompts (P3 to P7). As a reminder, prompt P3 is the most straightforward prompt, containing no additional information. Prompt P4 includes the location of the bug within the code snippet. Prompt P5 provides the range of locations that need to be fixed within the code. Prompt P6 inserts the review comment into the buggy code snippet. Lastly, prompt P7 combines the information from P5 and P6, including the range of fix locations and the review comment.

In Table \ref{table:main}, a darker color represents a higher value within the same row. The performance of the models generally improves from left to right for each prompt. Regarding the ECM column, both \codellama and \codegent consistently achieve the first and second highest results across all prompts. Notably, the best performance of \codellama attains a score of ECM as high as 72.97\%. \codefuse achieves the third-best performance in prompts P3 and P4, while \llama achieves the third-best performance in prompts P5 to P7. The last three prompts, which provide additional information in natural language, are more beneficial for \llama, as it exhibits better comprehension of such information than \codefuse. \incoder and \codetfp demonstrate the worst ECM values despite being primarily code language models. However, even with the lowest ECM scores, they can still fix over 62\% of the CR defects when prompted with P7.

\begin{figure}[htb]
\small
\begin{lstlisting}[language=Java,escapechar=|]
  public byte[] getData() {
|\gitdelete[4]{Object base = slice.getBase();}|
|\gitdelete[4]{checkState(base instanceof byte[], ...);}|
|\gitdelete[4]{byte[] bytes = (byte[]) base;}|
    |\cmtline{Should the above three lines actually happen in `if` branch? I don't see why we would need to get `base` if  the slice is not compact.}|
    if(slice.isCompact()) {
|\gitadd[6]{Object base = slice.getBase();}|
|\gitadd[6]{checkState(base instanceof byte[], ...);}|
|\gitadd[6]{return (byte[]) base;}|
|\gitdelete[6]{return bytes;}|
    }
    return slice.getBytes();
  }
\end{lstlisting}
\vskip -2mm
\caption{A buggy code snippet that was only fixed by \codellama with P7. \fixlocs is before line 2 and \fixloce is after line 10, they are omitted for simplicity.}\label{fig:code:fix-by-codellama}
\end{figure}

\Cref{fig:code:fix-by-codellama} presents an illustrative example where only \codellama, prompted with P7, successfully fixes the code defect. The reviewer’s comment clearly describes the defect's underlying cause and specifies the range of code lines that need to be modified. This defect requires multiple code modifications and involves complex information conveyed through natural language comments. This example showcases the distinct advantage of \codellama's advanced capabilities.

In the Code BLEU column, \codellama achieves the highest score, while \llama consistently secures the second position across all prompts instead of \codegent. Code BLEU is calculated based on the raw output, which includes natural language code comments. The overall trend observed in the Code BLEU scores aligns with the trend of the ECM, indicating the validity and reasonableness of using ECM as an evaluation metric for assessing defect repair capabilities. The differing rankings of the models regarding ECM and Code BLEU emphasize that Code BLEU may not be as suitable as ECM for evaluating code generation tasks. This conclusion aligns with the findings presented in \cite{bleu-bad}.

\codereviewer is pretrained with specially formatted code change data on a model base with 220 million parameters. However, it should be noted that this model lacks training in natural language, which influenced our decision to refrain from finetuning it with all prompts. Instead, we performed finetuning on CodeReviewer, utilizing its input-output data format and incorporating the review comment within the training process. This approach is similar to prompt P6, where the review comment is inserted into the buggy code snippet. The resulting ECM and Code BLEU scores for \codereviewer are 47.87\% and 0.8444, respectively. These scores are lower compared to the zero-shot learning performance of \gptt and all other models after finetuning. However, considering the smaller model size of \codereviewer, it remains competitive among all the models evaluated in this context.

\subsubsection{Detailed Study of the Defects in PD}

\newcommand{\desc}[1]{#1}
\newcommand{\pnum}[1]{#1\%}

\begin{table}[htb]
\caption{The lowest and hightest ECM of repairing each type of CR defects identified by PMD.}
\label{table:bugtype}
\begin{threeparttable}
\centering
\small
\begin{tabular*}{0.85\columnwidth}{l|r|r}
    \toprule
    \textbf{Short Description}  		            & \textbf{Lowest} 	            & \textbf{Highest}\\ 
    \midrule
    \desc{Unnecessary return} &\pnum{96.60} &\cellcolor{green} \pnum{100.00} \\
    \desc{Empty statement} &\pnum{86.80} & \pnum{97.20} \\
    \desc{Use isEmpty()} &\pnum{82.20} & \pnum{91.30} \\
    \desc{Simplified to while loop} &\pnum{81.80} & \pnum{90.90} \\
    \desc{Use existing BigDecimal instances} &\pnum{81.80} & \pnum{90.90} \\
    \desc{Add default in switch} &\pnum{80.80} & \pnum{90.70} \\
    \desc{Use String.indexOf(char)} &\pnum{78.40} & \pnum{94.60} \\
    \desc{Convert primitives to Strings directly} &\pnum{72.70} & \pnum{81.80} \\
    \desc{Literal first in comparison} &\pnum{71.70} & \pnum{82.70} \\
    \desc{Compare boolean literal} &\pnum{66.70} & \pnum{79.50} \\
    \desc{Use instanceof w/o null check} &\pnum{63.30} & \pnum{86.70} \\
    \desc{Nested if} &\pnum{57.30} & \pnum{72.40} \\
    \desc{Return boolean directly} &\pnum{52.30} & \pnum{76.10} \\
    \desc{Make properties in enum final} &\pnum{50.00} & \pnum{81.30} \\
    \desc{Use BigDecimal.valueOf()} &\pnum{45.50} &\cellcolor{green} \pnum{100.00} \\
    \desc{No instantiating Boolean object} &\pnum{45.50} & \pnum{90.90} \\
    \desc{DateFormatter is not thread safe} &\pnum{45.50} & \pnum{90.90} \\
    \desc{No instantiating String object} &\pnum{40.00} & \pnum{80.00} \\
    \desc{Use equals instead of ``==''} &\pnum{40.00} & \pnum{70.00} \\
    \desc{Append literal to String directly} &\pnum{38.50} & \pnum{92.30} \\
    \desc{Make final fields static} &\pnum{33.30} & \pnum{73.30} \\
    \desc{Inefficient String Buffer} &\pnum{31.00} & \pnum{48.30} \\
    \desc{Remove unused variables} &\pnum{30.00} &\cellcolor{green} \pnum{100.00} \\
    \desc{Use String.equalsIgnoreCase()} &\pnum{27.30} & \pnum{72.70} \\
    \desc{Use StringUtils.isBlank()} &\pnum{26.70} & \pnum{66.70} \\
    \desc{Use separate catch clause} &\pnum{23.50} & \pnum{58.80} \\
    \desc{Return value w/o storing} &\pnum{22.20} & \pnum{77.80} \\
    \desc{Use ``=='' to check null} &\pnum{9.10} & \pnum{54.50} \\
    \desc{Remove unused imports} &\cellcolor{gray!20}\pnum{0.00} &\cellcolor{green} \pnum{100.00} \\
    \desc{Catch exception when parsing double} &\cellcolor{gray!20}\pnum{0.00} & \pnum{6.70} \\
    \bottomrule
\end{tabular*}
\begin{tablenotes}[flushleft]\footnotesize
    \item The value is compared among all finetuned models under all prompts (\ie, P3$\sim$P7). The table is sorted in descending order of lowest ECM.
\end{tablenotes}
\end{threeparttable}
\end{table}


Our dataset has 30 different types of CR defects obtained from PMD checkers (PD). To compare the performance of each model across different prompts, we examined the ECM for each defect type. \Cref{table:bugtype} overviews the lowest and highest ECM values obtained among all the finetuning settings for each defect type. The first column includes a brief description of each defect. Some defects are relatively easier to repair as comments can explicitly describe them, or the code elements that need to be changed are easily identifiable. Notably, four defect types achieved a 100\% fix rate, which are highlighted in green in \cref{table:bugtype}.

Among the various types of defects, the last type is the most challenging to fix. As illustrated in Figure \cref{fig:code:parse-double}, the correct code fix involves enclosing the \code{Double.parseDouble} invocation within \code{try-catch} blocks. However, different code generation models may generate different exception messages in the repaired code, deviating from the expected output. Another complex defect type to fix, particularly with prompt P3, is the "Remove unused imports" issue. Without any indication of which import statements are unused, most models struggle to accurately identify and remove the target imports. For defects with lower scores, a potential approach to improvement is to gather additional high-quality data and perform further finetuning.

\begin{figure}[htb]
\small
\begin{lstlisting}[language=Java,escapechar=|]
|\gitadd[2]{try \{}|
    position.set(Double.parseDouble(values[index]));
|\gitadd[2]{\} catch(NumberFormatException e) \{}|
|\gitadd[4]{throw new  NumberFormatException("...");}|
|\gitadd[2]{\}}|
\end{lstlisting}
\vskip -2mm
\caption{A correct fix of ``catch exception when parsing double'' defect. The generated exception message in line 4 can be easily different.}\label{fig:code:parse-double}
\end{figure}

\subsection{Prompt Comparison (RQ2)}

\Cref{table:main} shows that all models perform the best in prompt P7, indicating that the combination of review comments and fix range significantly aids in repairing CR defects. The enhanced performance suggests that incorporating review comments and information about the fix range helps the models better understand and address the defects present in the code. On the other hand, the models perform the worst in prompt P3. This is because P3 only provides information that the code snippet is buggy without additional context or guidance. Without explicit details or cues from review comments or fix range, the models struggle to accurately identify and repair the CR defects in this prompt. The last row of the table, denoted as $\#{P7}$-$\#{P3}$, represents the improvement of each model from the lowest value (in P3) to the highest value (in P7). It is noteworthy that all models exhibit improvements in ECM that are uniformly greater than 20\%. This indicates that incorporating review comments and fix range information substantially enhances the models' ability to correct CR defects.

\begin{figure}[htb]
\small
\begin{lstlisting}[language=Java,escapechar=|]
  public byte[] getData(){
    if (slice.isCompact()) {
      return slice.getBytes();
    }
    return slice.getBytes();
  }
\end{lstlisting}
\vskip -2mm
\caption{The incorrect fixed code snippets generated by \codellama with prompt P5 and P6. The buggy code and the expected fixed code are shown in \cref{fig:code:fix-by-codellama}.}\label{fig:code:not-fix-by-codellama}
\end{figure}

\Cref{fig:code:not-fix-by-codellama} presents the incorrect fixed code snippets generated by \codellama with prompts P5 and P6. In contrast, \codellama with prompt P7 can successfully repair the defect, as shown in \cref{fig:code:fix-by-codellama}.
Only given the review comment or the fix range, \codellama cannot come up with the complete patch and only remove the three lines above the \code{if} statement and change line 3.

\begin{table}[htb]
    \caption{The improvement between adjacent prompts of each model.}
    \label{table:prompt-improve}
    \centering
    \small
\begin{tabular*}{\columnwidth}{wc{0.7cm}|wc{0.6cm}wc{0.7cm}wc{0.6cm}wc{0.6cm}wc{0.7cm}wc{1cm}|c}
    \toprule
    & \textbf{\scriptsize \incoder} & \textbf{\scriptsize\codetfp} & \textbf{\scriptsize\codefuse} & \textbf{\scriptsize\llama} & \textbf{\scriptsize\codegent} & \textbf{\scriptsize\codellama} & \textbf{\scriptsize Average} \\
    \midrule
    $\#_{P4}$-$\#_{P3}$ & 5.19\%	&5.26\%	    &5.45\%	    &6.58\%	    &5.55\%	    &6.61\%	    &5.77\% \\
    $\#_{P5}$-$\#_{P4}$ & 7.55\%	&7.03\%	    &6.77\%	    &7.48\%	    &8.84\%	    &8.49\%	    &7.69\%  \\
    $\#_{P6}$-$\#_{P5}$ & 5.20\%	&3.97\%	    &3.65\%	    &3.65\%	    &1.48\%	    &3.42\%	    &3.56\% \\
    $\#_{P7}$-$\#_{P6}$ & 4.71\%	&5.38\%	    &5.64\%	    &5.93\%	    &6.39\%	    &6.58\%	    &5.77\% \\
    \bottomrule
\end{tabular*}
\end{table}

To analyze the improvement in ECM between adjacent prompts, we calculated the value of $\#{Pn+1}$-$\#{Pn}$ in \cref{table:prompt-improve}. The results provide insights into the impact of incorporating different cues in each prompt. From P3 to P4, the introduction of bug location information leads to an average improvement in ECM of 5.77\%. This improvement suggests that identifying the location of the defect provides valuable context for the models, assisting them in better understanding and addressing the defects. From P4 to P5, replacing defect location with fix range information results in the highest improvement in ECM, with an average increase of 7.69\%. This improvement highlights the significance of fix range information, because it represents the golden locations for repairing defects. The fact that the defect location and the fix locations are often different emphasizes the importance of incorporating the fix range, as it provides more accurate and specific guidance for repairing the defects. In P6, the inclusion of reviewer comments leads to an average improvement of 3.56\% in ECM compared to P5. This improvement demonstrates the value of incorporating human feedback and insights in the form of reviewer comments, enabling the models to understand better and address the defects present in the code. Finally, in P7, the combination of the two powerful cues, \ie reviewer comments and fix range, further improves the ECM by an additional 5.77\%. This improvement signifies the synergistic effect of incorporating multiple cues, resulting in a more effective and accurate repair process.

Overall, the analysis of the improvements in ECM between adjacent prompts highlights the incremental benefits of incorporating different cues. Each prompt introduces a new cue or modifies an existing one, leading to a progressive enhancement in the models' ability to repair CR defects. The results emphasize the importance of considering multiple cues and incorporating diverse information sources to improve models' performance in code repair tasks.

\subsection{Model Size Comparison (RQ3)}

\begin{figure}[htb]
    \centering
    \small
    \begin{tikzpicture}[baseline]
        \pgfplotsset{
            height = 4cm,
            width  = 8cm,
        }
        \begin{axis}[
            axis x line*=bottom,
            xtick={0.22, 0.77, 2, 6, 16},
            xtick distance=1,
            xticklabel style={
                rotate=45, anchor=east,
                text height= 2ex,
            },
            ylabel near ticks, 
            yticklabel pos=left,
            ymin=0.89, ymax=0.92, 
            ylabel=Code BLEU,
        ]
        \addplot[style={cyan,mark=star}]
            coordinates {
                (0.22, 0.8936)
                (0.77, 0.9029)
                (2, 0.9035)
                (6, 0.9011)
                (16, 0.9116)
            };
        \label{cb}
        \end{axis}
        \begin{axis}[
            axis x line*=bottom,
            xlabel=Model size (B),
            xtick={0.22, 0.77, 2, 6, 16},
            xtick distance=1,
            xticklabel style={
                rotate=45, anchor=east,           
                text height= 2ex,
            },
            ylabel near ticks, 
            yticklabel pos=right,
            ytick={52.00, 57.00, 62.00, 67.00},
            ymin=52.00, ymax=67.00,
            ylabel=ECM (\%), 
            legend style={
                legend columns=2,
                draw=none, 
                /tikz/every even column/.append style={column sep=0.3cm},
                legend to name=sizecomp
            },
        ]
        \addlegendimage{/pgfplots/refstyle=cb}\addlegendentry{Code BLEU}
        \addplot[style={orange,mark=triangle}]
            coordinates {
                (0.22, 54.00)
                (0.77, 60.81)
                (2, 62.97)
                (6, 63.03)
                (16, 66.29)
            };
        \addlegendentry{ECM}
        \end{axis}
    \end{tikzpicture}
    \ref{sizecomp}
    \caption{The performance comparison of \codetfp at different sizes (0.22B, 0.77B,2B, 6B, 16B).}
    \label{fig:diffsize}
\end{figure}
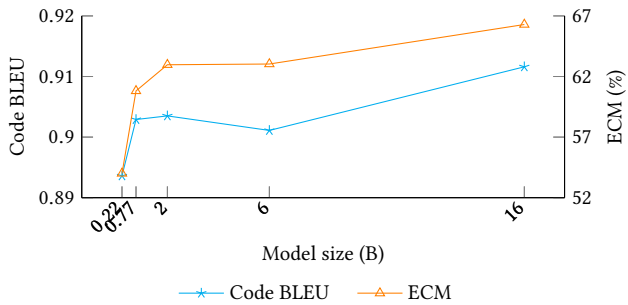

To examine the impacts of model size, we conducted finetuning experiments on five models from \codetfp, which offers the most extensive range of candidate sizes available, spanning from 220 million parameters to 16 billion parameters. \Cref{fig:diffsize} demonstrates the trends of the ECM and Code BLEU scores in \codetfp. Overall, both ECM and Code BLEU scores gradually increased, except for a slight decrease in Code BLEU from 2 billion parameters (0.9035) to 6 billion parameters (0.9011). The most significant improvement in both ECM and Code BLEU scores is observed when transitioning from the 220 million to the 770 million parameter model. As the model size increases further, the rate of improvement gradually slows down, particularly from the 2 billion parameter model to the 6 billion parameter model.

Considering both efficiency and effectiveness, the 6 billion or 7 billion parameter models are the most practical choices compared to smaller or larger models. Additionally, we conducted finetuning and validation on the 13 billion parameter models of \codefuse, \llama, and \codellama using prompt P7. Notably, we observed similar growth trends in their performance. The ECM showed a slight increase for \codefuse (2.2\%) and \llama (1.87\%), while a minor decrease of 0.29\% was noticed for \codellama. These findings further support that the intermediate-sized models are a practical and effective choice in our scenario.

\subsection{Impacts of Datasets (RQ4)}
\newcommand{\clr}{Code\-LLaMA$_{RD}$\xspace} 
\newcommand{\rt}{RD$_t$\xspace} 
\newcommand{\clp}{Code\-LLaMA$_{PD}$\xspace} 
\newcommand{\pt}{PD$_t$\xspace} 

In order to evaluate the different impacts of datasets from human reviewers and automated checkers, we conducted separate finetuning experiments on \codellama using each dataset. The model finetuned on the RD and PD is denoted as \clr and \clp, respectively. We then cross-validated the two resulting models by evaluating them on both the RD test set, denoted as $RD_t$, and the PD test set, denoted as $PD_t$. \Cref{table:diffdata} presents the ECM scores. The findings indicate that finetuning and testing on homologous datasets yield the best results. For example, the model \clr can fix more defects in $RD_t$ compared to $PD_t$, even though defects in the PD dataset may be more prevalent. Conversely, the scores of testing on the heterologous data are approximately decreased by half. Furthermore, the combined scores of the two models on each test set show only marginal improvements, as indicated in the last row of \cref{table:diffdata}.

\begin{table}[htb]
    \begin{threeparttable}
    \caption{Performance of finetuning \codellama with one dataset and validate on another.}
    \label{table:diffdata}
    \begin{tabular}{@{}p{\columnwidth}@{}}
    \centering
        \small
        \begin{tabular}{c|cc}
        \toprule
        \multirow{2}{*}{\textbf{Finetuned Model}}  & \multicolumn{2}{c}{\textbf{ECM}}\\ 
        & \rt & \pt \\
        \midrule
        \clr    & 65.10\% & 44.18\% \\
        \clp    & 36.31\% & 80.65\% \\
        \midrule
        \textbf{Union of Fix} & 67.02\% & 82.41\% \\
        \bottomrule
        \end{tabular}
    \end{tabular}
    \begin{tablenotes}[flushleft]\footnotesize
        \item We use prompt P7 in this comparison.
        The first column shows the two \codellama models finetuned on different datasets.
        The last row is calculated by combing the correct fixes generated by the two finetuned models.    \end{tablenotes}
\end{threeparttable}
\end{table}

The above results signify the importance of training and evaluating models on the appropriate datasets to achieve optimal performance and highlight the necessity of the PD and RD datasets in the finetuning process. In other words, finetuning models on diverse datasets that represent human and automated perspectives allows for a more comprehensive and practical training procedure.

\section{Threats}\label{sec:threats}
We acknowledge that there may be certain threats to the validity of our findings. Our study primarily focused on the performance of 6$\sim$7B LLMs, but it is important to note that there are numerous other models of varying scales available in the open source community. In our research, we emphasized the significance of both inference efficiency and effectiveness in order to ensure smooth integration within the CI/CD process. The 6-7B LLMs demonstrate a favorable balance between efficiency and effectiveness, able to infer within seconds while possessing strong capabilities compared to smaller models. To validate our perspective, we conducted an experiment comparing different scales of the \codetfp model. The results reinforced our belief that the 6-7B model achieves an optimal balance between efficiency and effectiveness.

Regarding data collection, it is essential to acknowledge potential biases that may influence the performance of LLMs. Data collection is a continual challenge in research, and we took manual efforts to ensure the quality of the dataset. Furthermore, we observed consistent trends in our results when comparing different models, prompts, and datasets, which suggests that the dataset meets the necessary standards.

\section{Related Work}\label{sec:related}

Automated program repair has been extensively studied. Existing APR tools typically involve three main steps: bug localization \cite{fl-survey-2016}, patch generation, and patch validation. Based on the techniques employed in the patch generation step, APR tools can be classified into several types \cite{apr-survey-2019-tse,apr-survey-2023-huang,apr-survey-2023-learning}: \emph{search-based}, \emph{constraint-based}, \emph{template-based}, and \emph{learning-based}.

\subsection{Conventional APR Approaches}
\emph{Search-based} methods aim to explore the vast space of potential patches by generating code variations at the most probable bug locations. Techniques such as evolutionary computation \cite{apr-ec-2010} and genetic programming \cite{apr-genprog-tse, apr-gp-icsm-2013, apr-gp-asplos-2013, apr-gp-icse-2009} are commonly employed to mutate the code at these bug locations. The search efficiency of search-based methods remains a critical challenge. The exploration of the patch space can be computationally expensive and time-consuming. Improving the efficiency of search-based methods is an ongoing focus of research in APR, as enhancing the effectiveness and practicality of these approaches is essential for their widespread adoption.

\emph{Constraint-based} methods transform the patch generation problem into a constraint solving problem and leverage specifications to exclude infeasible patches. These methods aim to overcome the challenges of search-based approaches by providing a more structured and guided approach to generating patches. Nopol \cite{apr-nopol} explicitly targets conditional bugs and utilizes tests as specifications to guide the patch generation process. SemFix \cite{apr-semfix} uses symbolic execution to generate repair constraints for fixing single-line bugs. Angelix \cite{apr-angelix}, on the other hand, introduces the concept of angelic forest as a lightweight constraint representation and proposes a technique to fix multi-line bugs using these constraints. \emph{Constraint-based} methods encounter challenges when they lack appropriate constraint specifications for specific types of bugs.

\emph{Template-based} approaches utilize predefined patch templates or patterns to generate patches for code defects. These templates can be extracted manually or automatically, providing a structured and guided approach to patch generation. For instance, Kim et al. \cite{apr-template-icse-13} manually reviewed patches and abstracted ten templates, forming the PAR tool's basis. Relifix \cite{apr-relifix-icse} employed syntax-level transformation rules manually extracted from version histories as templates. Alternatively, automatic techniques have been developed to mine templates from project histories. HDRepair \cite{apr-hdrepair} utilized graph mining to automatically abstract fix patterns, while Genesis \cite{apr-genesis} extracted transformation rules from sampled code transformations. Recent studies \cite{apr-survey-2019-tse} have shown that template-based methods outperform search-based and constraint-based methods in terms of effectiveness. However, one limitation of template-based tools is their rigidity, as they can only fix specific types of bugs that align with the predefined templates. This restricts their applicability and adaptability to a broader range of defects.

\subsection{Intelligent APR Approaches}

\emph{Learning-based} APR tools utilize bug repair samples to acquire empirical knowledge and generate patches. These data-driven repair solutions leverage machine learning techniques to learn from the available data. One approach is to treat the APR problem as a translation task, where the model learns to translate from buggy code to fixed code. Tufano et al. \cite{apr-tufano-nmt-2019} employed a neural machine translation (NMT) model to train on pairs of buggy and fixed methods. The model learned the mapping between the two, enabling it to generate patches. Although learning-based APR tools have shown promise, they may have limitations in fully understanding natural language comments. Improving their capability to comprehend and effectively utilize review comments can enhance the quality and accuracy of generated patches.

Benefiting from the rapid development of LLMs, APR approaches also utilized LLMs to repair bugs.
Circle \cite{apr-circle-issta-2022} designed prompts that include context, buggy code, and fixed code, using natural language sentences to separate different parts. CoCoNut \cite{apr-coconut-issta-2020}, on the other hand, utilizes two encoders to encode the buggy code and context separately. 
Different from this paper, the two approaches aim at general bugs.
InferFix \cite{inferfix} employs LLMs to repair bugs detected by an automated checker (Infer \cite{infer}), including null pointer dereference, resource leak, and thread safety violations. They fine-tune Codex \cite{codex-2021} using prompts that contain bug type, location, and the context of the buggy method. They also collect a dataset of previous fixes and query similar fixes to include in the prompt. However, InferFix only focuses on a few types of bugs identified by automated checkers and does not consider reviewer comments.

\section{Conclusion}\label{sec:conclusion}

In this paper, we investigate the LLMs in repairing CR defects, focusing on the valuable insights provided by accompanying review comments.
Our systematic experiment compares 9 models in zero-shot learning and finetuning scenarios, revealing that \codellama performs the best with a repair rate of up to 72.97\% for CR defects.
Additionally, we design 7 prompts containing different useful information, highlighting the effectiveness of review comments and fix ranges.
By comparing the performance of different scales of \codetfp,
the 6$\sim$7B models are proved to be the practical and effective choice in repairing CR defects.
Furthermore, we also find the poor transferability of learned abilities from different datasets, emphasizing the need for finetuning with diverse datasets of comparable size. Our paper provides valuable insights for leveraging LLMs in CR defect repair to enhance software quality and efficiency in the CI/CD pipeline.

\bibliographystyle{ACM-Reference-Format}
\bibliography{sample-base}


\begin{thebibliography}{38}


\ifx \showCODEN    \undefined \def \showCODEN     #1{\unskip}     \fi
\ifx \showDOI      \undefined \def \showDOI       #1{#1}\fi
\ifx \showISBNx    \undefined \def \showISBNx     #1{\unskip}     \fi
\ifx \showISBNxiii \undefined \def \showISBNxiii  #1{\unskip}     \fi
\ifx \showISSN     \undefined \def \showISSN      #1{\unskip}     \fi
\ifx \showLCCN     \undefined \def \showLCCN      #1{\unskip}     \fi
\ifx \shownote     \undefined \def \shownote      #1{#1}          \fi
\ifx \showarticletitle \undefined \def \showarticletitle #1{#1}   \fi
\ifx \showURL      \undefined \def \showURL       {\relax}        \fi
\providecommand\bibfield[2]{#2}
\providecommand\bibinfo[2]{#2}
\providecommand\natexlab[1]{#1}
\providecommand\showeprint[2][]{arXiv:#2}

\bibitem[Chen et~al\mbox{.}(2021)]%
        {codex-2021}
\bibfield{author}{\bibinfo{person}{Mark Chen}, \bibinfo{person}{Jerry Tworek},
  \bibinfo{person}{Heewoo Jun}, \bibinfo{person}{Qiming Yuan},
  \bibinfo{person}{Henrique~Pond{\'{e}} de Oliveira~Pinto}, {et~al\mbox{.}}}
  \bibinfo{year}{2021}\natexlab{}.
\newblock \bibinfo{title}{Evaluating Large Language Models Trained on Code}.
\newblock
\newblock
\showeprint[arXiv]{2107.03374}
\urldef\tempurl%
\url{https://arxiv.org/abs/2107.03374}
\showURL{%
\tempurl}


\bibitem[Devlin et~al\mbox{.}(2019)]%
        {bert}
\bibfield{author}{\bibinfo{person}{Jacob Devlin}, \bibinfo{person}{Ming-Wei
  Chang}, \bibinfo{person}{Kenton Lee}, {and} \bibinfo{person}{Kristina
  Toutanova}.} \bibinfo{year}{2019}\natexlab{}.
\newblock \showarticletitle{BERT: Pre-training of Deep Bidirectional
  Transformers for Language Understanding}.
\newblock  (\bibinfo{year}{2019}).
\newblock
\showeprint[arxiv]{1810.04805}~[cs.CL]


\bibitem[Evtikhiev et~al\mbox{.}(2023)]%
        {bleu-bad}
\bibfield{author}{\bibinfo{person}{Mikhail Evtikhiev}, \bibinfo{person}{Egor
  Bogomolov}, \bibinfo{person}{Yaroslav Sokolov}, {and}
  \bibinfo{person}{Timofey Bryksin}.} \bibinfo{year}{2023}\natexlab{}.
\newblock \showarticletitle{Out of the BLEU: How Should We Assess Quality of
  the Code Generation Models?}
\newblock \bibinfo{journal}{\emph{J. Syst. Softw.}} \bibinfo{volume}{203},
  \bibinfo{number}{C} (\bibinfo{date}{jul} \bibinfo{year}{2023}),
  \bibinfo{numpages}{17}~pages.
\newblock
\showISSN{0164-1212}
\urldef\tempurl%
\url{https://doi.org/10.1016/j.jss.2023.111741}
\showDOI{\tempurl}


\bibitem[Fried et~al\mbox{.}(2022)]%
        {incoder-23}
\bibfield{author}{\bibinfo{person}{Daniel Fried}, \bibinfo{person}{Armen
  Aghajanyan}, \bibinfo{person}{Jessy Lin}, \bibinfo{person}{Sida~I. Wang},
  \bibinfo{person}{Eric Wallace}, \bibinfo{person}{Freda Shi},
  \bibinfo{person}{Ruiqi Zhong}, \bibinfo{person}{Wen tau Yih},
  \bibinfo{person}{Luke Zettlemoyer}, {and} \bibinfo{person}{Mike Lewis}.}
  \bibinfo{year}{2022}\natexlab{}.
\newblock \showarticletitle{InCoder: A Generative Model for Code Infilling and
  Synthesis}.
\newblock \bibinfo{journal}{\emph{ArXiv}}  \bibinfo{volume}{abs/2204.05999}
  (\bibinfo{year}{2022}).
\newblock
\urldef\tempurl%
\url{https://api.semanticscholar.org/CorpusID:248157108}
\showURL{%
\tempurl}


\bibitem[Gazzola et~al\mbox{.}(2019)]%
        {apr-survey-2019-tse}
\bibfield{author}{\bibinfo{person}{Luca Gazzola}, \bibinfo{person}{Daniela
  Micucci}, {and} \bibinfo{person}{Leonardo Mariani}.}
  \bibinfo{year}{2019}\natexlab{}.
\newblock \showarticletitle{Automatic Software Repair: A Survey}.
\newblock \bibinfo{journal}{\emph{IEEE Transactions on Software Engineering}}
  \bibinfo{volume}{45}, \bibinfo{number}{1} (\bibinfo{year}{2019}),
  \bibinfo{pages}{34--67}.
\newblock
\urldef\tempurl%
\url{https://doi.org/10.1109/TSE.2017.2755013}
\showDOI{\tempurl}


\bibitem[Huang et~al\mbox{.}(2023)]%
        {apr-survey-2023-huang}
\bibfield{author}{\bibinfo{person}{Kai Huang}, \bibinfo{person}{Zhengzi Xu},
  \bibinfo{person}{Su Yang}, \bibinfo{person}{Hongyu Sun},
  \bibinfo{person}{Xuejun Li}, \bibinfo{person}{Zheng Yan}, {and}
  \bibinfo{person}{Yuqing Zhang}.} \bibinfo{year}{2023}\natexlab{}.
\newblock \bibinfo{title}{A Survey on Automated Program Repair Techniques}.
\newblock
\newblock
\showeprint[arxiv]{2303.18184}~[cs.SE]


\bibitem[Inc.(2023a)]%
        {codefuse}
\bibfield{author}{\bibinfo{person}{AntGroup Inc.}}
  \bibinfo{year}{2023}\natexlab{a}.
\newblock \bibinfo{booktitle}{\emph{{CodeFuse}}}.
\newblock
\urldef\tempurl%
\url{https://github.com/codefuse-ai}
\showURL{%
\tempurl}
\newblock
\shownote{Accessed: 2023-09-22}.


\bibitem[Inc.(2023b)]%
        {infer}
\bibfield{author}{\bibinfo{person}{Facebook Inc.}}
  \bibinfo{year}{2023}\natexlab{b}.
\newblock \bibinfo{booktitle}{\emph{{Infer}}}.
\newblock
\urldef\tempurl%
\url{https://github.com/facebook/infer}
\showURL{%
\tempurl}
\newblock
\shownote{Accessed: 2023-09-22}.


\bibitem[Inc.(2023c)]%
        {chatgpt}
\bibfield{author}{\bibinfo{person}{OpenAI Inc.}}
  \bibinfo{year}{2023}\natexlab{c}.
\newblock \bibinfo{booktitle}{\emph{{ChatGPT}}}.
\newblock
\urldef\tempurl%
\url{https://openai.com/blog/chatgpt}
\showURL{%
\tempurl}
\newblock
\shownote{Accessed: 2023-09-22}.


\bibitem[Jin et~al\mbox{.}(2023)]%
        {inferfix}
\bibfield{author}{\bibinfo{person}{Matthew Jin}, \bibinfo{person}{Syed
  Shahriar}, \bibinfo{person}{Michele Tufano}, \bibinfo{person}{Xin Shi},
  \bibinfo{person}{Shuai Lu}, \bibinfo{person}{Neel Sundaresan}, {and}
  \bibinfo{person}{Alexey Svyatkovskiy}.} \bibinfo{year}{2023}\natexlab{}.
\newblock \bibinfo{title}{InferFix: End-to-End Program Repair with LLMs}.
\newblock
\newblock
\showeprint[arxiv]{2303.07263}~[cs.SE]


\bibitem[Kim et~al\mbox{.}(2013)]%
        {apr-template-icse-13}
\bibfield{author}{\bibinfo{person}{Dongsun Kim}, \bibinfo{person}{Jaechang
  Nam}, \bibinfo{person}{Jaewoo Song}, {and} \bibinfo{person}{Sunghun Kim}.}
  \bibinfo{year}{2013}\natexlab{}.
\newblock \showarticletitle{Automatic Patch Generation Learned from
  Human-Written Patches}. In \bibinfo{booktitle}{\emph{Proceedings of the 2013
  International Conference on Software Engineering}}
  \emph{(\bibinfo{series}{ICSE '13})}. \bibinfo{pages}{802–811}.
\newblock
\showISBNx{9781467330763}


\bibitem[Le et~al\mbox{.}(2016)]%
        {apr-hdrepair}
\bibfield{author}{\bibinfo{person}{Xuan Bach~D. Le}, \bibinfo{person}{David
  Lo}, {and} \bibinfo{person}{Claire Le~Goues}.}
  \bibinfo{year}{2016}\natexlab{}.
\newblock \showarticletitle{History Driven Program Repair}. In
  \bibinfo{booktitle}{\emph{2016 IEEE 23rd International Conference on Software
  Analysis, Evolution, and Reengineering (SANER)}}, Vol.~\bibinfo{volume}{1}.
  \bibinfo{pages}{213--224}.
\newblock
\urldef\tempurl%
\url{https://doi.org/10.1109/SANER.2016.76}
\showDOI{\tempurl}


\bibitem[Le~Goues et~al\mbox{.}(2012)]%
        {apr-genprog-tse}
\bibfield{author}{\bibinfo{person}{Claire Le~Goues}, \bibinfo{person}{ThanhVu
  Nguyen}, \bibinfo{person}{Stephanie Forrest}, {and} \bibinfo{person}{Westley
  Weimer}.} \bibinfo{year}{2012}\natexlab{}.
\newblock \showarticletitle{GenProg: A Generic Method for Automatic Software
  Repair}.
\newblock \bibinfo{journal}{\emph{IEEE Transactions on Software Engineering}}
  \bibinfo{volume}{38}, \bibinfo{number}{1} (\bibinfo{year}{2012}),
  \bibinfo{pages}{54--72}.
\newblock
\urldef\tempurl%
\url{https://doi.org/10.1109/TSE.2011.104}
\showDOI{\tempurl}


\bibitem[Li et~al\mbox{.}(2022)]%
        {codereviewer-22}
\bibfield{author}{\bibinfo{person}{Zhiyu Li}, \bibinfo{person}{Shuai Lu},
  \bibinfo{person}{Daya Guo}, \bibinfo{person}{Nan Duan},
  \bibinfo{person}{Shailesh Jannu}, \bibinfo{person}{Grant Jenks},
  \bibinfo{person}{Deep Majumder}, \bibinfo{person}{Jared Green},
  \bibinfo{person}{Alexey Svyatkovskiy}, \bibinfo{person}{Shengyu Fu}, {and}
  \bibinfo{person}{Neel Sundaresan}.} \bibinfo{year}{2022}\natexlab{}.
\newblock \showarticletitle{Automating Code Review Activities by Large-Scale
  Pre-Training}. In \bibinfo{booktitle}{\emph{Proceedings of the 30th ACM Joint
  European Software Engineering Conference and Symposium on the Foundations of
  Software Engineering}} \emph{(\bibinfo{series}{ESEC/FSE 2022})}.
  \bibinfo{pages}{1035–1047}.
\newblock
\showISBNx{9781450394130}
\urldef\tempurl%
\url{https://doi.org/10.1145/3540250.3549081}
\showDOI{\tempurl}


\bibitem[Long et~al\mbox{.}(2017)]%
        {apr-genesis}
\bibfield{author}{\bibinfo{person}{Fan Long}, \bibinfo{person}{Peter Amidon},
  {and} \bibinfo{person}{Martin Rinard}.} \bibinfo{year}{2017}\natexlab{}.
\newblock \showarticletitle{Automatic Inference of Code Transforms for Patch
  Generation}. In \bibinfo{booktitle}{\emph{Proceedings of the 2017 11th Joint
  Meeting on Foundations of Software Engineering}}
  \emph{(\bibinfo{series}{ESEC/FSE 2017})}. \bibinfo{pages}{727–739}.
\newblock
\showISBNx{9781450351058}
\urldef\tempurl%
\url{https://doi.org/10.1145/3106237.3106253}
\showDOI{\tempurl}


\bibitem[Lutellier et~al\mbox{.}(2020)]%
        {apr-coconut-issta-2020}
\bibfield{author}{\bibinfo{person}{Thibaud Lutellier},
  \bibinfo{person}{Hung~Viet Pham}, \bibinfo{person}{Lawrence Pang},
  \bibinfo{person}{Yitong Li}, \bibinfo{person}{Moshi Wei}, {and}
  \bibinfo{person}{Lin Tan}.} \bibinfo{year}{2020}\natexlab{}.
\newblock \showarticletitle{CoCoNuT: Combining Context-Aware Neural Translation
  Models Using Ensemble for Program Repair}. In
  \bibinfo{booktitle}{\emph{Proceedings of the 29th ACM SIGSOFT International
  Symposium on Software Testing and Analysis}} \emph{(\bibinfo{series}{ISSTA
  2020})}. \bibinfo{pages}{101–114}.
\newblock
\showISBNx{9781450380089}
\urldef\tempurl%
\url{https://doi.org/10.1145/3395363.3397369}
\showDOI{\tempurl}


\bibitem[Mechtaev et~al\mbox{.}(2016)]%
        {apr-angelix}
\bibfield{author}{\bibinfo{person}{Sergey Mechtaev}, \bibinfo{person}{Jooyong
  Yi}, {and} \bibinfo{person}{Abhik Roychoudhury}.}
  \bibinfo{year}{2016}\natexlab{}.
\newblock \showarticletitle{Angelix: Scalable Multiline Program Patch Synthesis
  via Symbolic Analysis}. In \bibinfo{booktitle}{\emph{Proceedings of the 38th
  International Conference on Software Engineering}}
  \emph{(\bibinfo{series}{ICSE '16})}. \bibinfo{pages}{691–701}.
\newblock
\showISBNx{9781450339001}
\urldef\tempurl%
\url{https://doi.org/10.1145/2884781.2884807}
\showDOI{\tempurl}


\bibitem[Nguyen et~al\mbox{.}(2013)]%
        {apr-semfix}
\bibfield{author}{\bibinfo{person}{Hoang Duong~Thien Nguyen},
  \bibinfo{person}{Dawei Qi}, \bibinfo{person}{Abhik Roychoudhury}, {and}
  \bibinfo{person}{Satish Chandra}.} \bibinfo{year}{2013}\natexlab{}.
\newblock \showarticletitle{SemFix: Program Repair via Semantic Analysis}. In
  \bibinfo{booktitle}{\emph{Proceedings of the 2013 International Conference on
  Software Engineering}} \emph{(\bibinfo{series}{ICSE '13})}.
  \bibinfo{pages}{772–781}.
\newblock
\showISBNx{9781467330763}


\bibitem[Nijkamp et~al\mbox{.}(2023)]%
        {codegen2-23}
\bibfield{author}{\bibinfo{person}{Erik Nijkamp}, \bibinfo{person}{Hiroaki
  Hayashi}, \bibinfo{person}{Caiming Xiong}, \bibinfo{person}{Silvio Savarese},
  {and} \bibinfo{person}{Yingbo Zhou}.} \bibinfo{year}{2023}\natexlab{}.
\newblock \bibinfo{title}{CodeGen2: Lessons for Training LLMs on Programming
  and Natural Languages}.
\newblock
\newblock
\showeprint[arxiv]{2305.02309}~[cs.LG]


\bibitem[PMD(2023)]%
        {pmd}
\bibfield{author}{\bibinfo{person}{PMD}.} \bibinfo{year}{2023}\natexlab{}.
\newblock \bibinfo{booktitle}{\emph{{PMD}}}.
\newblock
\urldef\tempurl%
\url{https://pmd.github.io}
\showURL{%
\tempurl}
\newblock
\shownote{Accessed: 2023-09-22}.


\bibitem[Qi et~al\mbox{.}(2013)]%
        {apr-gp-icsm-2013}
\bibfield{author}{\bibinfo{person}{Yuhua Qi}, \bibinfo{person}{Xiaoguang Mao},
  {and} \bibinfo{person}{Yan Lei}.} \bibinfo{year}{2013}\natexlab{}.
\newblock \showarticletitle{Efficient Automated Program Repair through
  Fault-Recorded Testing Prioritization}. In \bibinfo{booktitle}{\emph{2013
  IEEE International Conference on Software Maintenance}}.
  \bibinfo{pages}{180--189}.
\newblock
\urldef\tempurl%
\url{https://doi.org/10.1109/ICSM.2013.29}
\showDOI{\tempurl}


\bibitem[Raffel et~al\mbox{.}(2020)]%
        {t5}
\bibfield{author}{\bibinfo{person}{Colin Raffel}, \bibinfo{person}{Noam
  Shazeer}, \bibinfo{person}{Adam Roberts}, \bibinfo{person}{Katherine Lee},
  \bibinfo{person}{Sharan Narang}, \bibinfo{person}{Michael Matena},
  \bibinfo{person}{Yanqi Zhou}, \bibinfo{person}{Wei Li}, {and}
  \bibinfo{person}{Peter~J. Liu}.} \bibinfo{year}{2020}\natexlab{}.
\newblock \showarticletitle{Exploring the Limits of Transfer Learning with a
  Unified Text-to-Text Transformer}.
\newblock \bibinfo{journal}{\emph{J. Mach. Learn. Res.}} \bibinfo{volume}{21},
  \bibinfo{number}{1}, Article \bibinfo{articleno}{140} (\bibinfo{date}{jan}
  \bibinfo{year}{2020}), \bibinfo{numpages}{67}~pages.
\newblock
\showISSN{1532-4435}


\bibitem[Ren et~al\mbox{.}(2020)]%
        {code-bleu}
\bibfield{author}{\bibinfo{person}{Shuo Ren}, \bibinfo{person}{Daya Guo},
  \bibinfo{person}{Shuai Lu}, \bibinfo{person}{Long Zhou},
  \bibinfo{person}{Shujie Liu}, \bibinfo{person}{Duyu Tang},
  \bibinfo{person}{M. Zhou}, \bibinfo{person}{Ambrosio Blanco}, {and}
  \bibinfo{person}{Shuai Ma}.} \bibinfo{year}{2020}\natexlab{}.
\newblock \showarticletitle{CodeBLEU: a Method for Automatic Evaluation of Code
  Synthesis}.
\newblock \bibinfo{journal}{\emph{ArXiv}}  \bibinfo{volume}{abs/2009.10297}
  (\bibinfo{year}{2020}).
\newblock
\urldef\tempurl%
\url{https://api.semanticscholar.org/CorpusID:221836101}
\showURL{%
\tempurl}


\bibitem[Rozière et~al\mbox{.}(2023)]%
        {codellama-23}
\bibfield{author}{\bibinfo{person}{Baptiste Rozière}, \bibinfo{person}{Jonas
  Gehring}, \bibinfo{person}{Fabian Gloeckle}, \bibinfo{person}{Sten Sootla},
  \bibinfo{person}{Itai Gat}, \bibinfo{person}{Xiaoqing~Ellen Tan},
  \bibinfo{person}{Yossi Adi}, \bibinfo{person}{Jingyu Liu},
  \bibinfo{person}{Tal Remez}, \bibinfo{person}{Jérémy Rapin},
  \bibinfo{person}{Artyom Kozhevnikov}, \bibinfo{person}{Ivan Evtimov},
  \bibinfo{person}{Joanna Bitton}, \bibinfo{person}{Manish Bhatt},
  \bibinfo{person}{Cristian~Canton Ferrer}, \bibinfo{person}{Aaron
  Grattafiori}, \bibinfo{person}{Wenhan Xiong}, \bibinfo{person}{Alexandre
  Défossez}, \bibinfo{person}{Jade Copet}, \bibinfo{person}{Faisal Azhar},
  \bibinfo{person}{Hugo Touvron}, \bibinfo{person}{Louis Martin},
  \bibinfo{person}{Nicolas Usunier}, \bibinfo{person}{Thomas Scialom}, {and}
  \bibinfo{person}{Gabriel Synnaeve}.} \bibinfo{year}{2023}\natexlab{}.
\newblock \bibinfo{title}{Code Llama: Open Foundation Models for Code}.
\newblock
\newblock
\showeprint[arxiv]{2308.12950}~[cs.CL]


\bibitem[Schulte et~al\mbox{.}(2013)]%
        {apr-gp-asplos-2013}
\bibfield{author}{\bibinfo{person}{Eric Schulte}, \bibinfo{person}{Jonathan
  DiLorenzo}, \bibinfo{person}{Westley Weimer}, {and}
  \bibinfo{person}{Stephanie Forrest}.} \bibinfo{year}{2013}\natexlab{}.
\newblock \showarticletitle{Automated Repair of Binary and Assembly Programs
  for Cooperating Embedded Devices}. In \bibinfo{booktitle}{\emph{Proceedings
  of the Eighteenth International Conference on Architectural Support for
  Programming Languages and Operating Systems}} \emph{(\bibinfo{series}{ASPLOS
  '13})}. \bibinfo{pages}{317–328}.
\newblock
\showISBNx{9781450318709}
\urldef\tempurl%
\url{https://doi.org/10.1145/2451116.2451151}
\showDOI{\tempurl}


\bibitem[Schulte et~al\mbox{.}(2010)]%
        {apr-ec-2010}
\bibfield{author}{\bibinfo{person}{Eric Schulte}, \bibinfo{person}{Stephanie
  Forrest}, {and} \bibinfo{person}{Westley Weimer}.}
  \bibinfo{year}{2010}\natexlab{}.
\newblock \showarticletitle{Automated Program Repair through the Evolution of
  Assembly Code}. In \bibinfo{booktitle}{\emph{Proceedings of the 25th IEEE/ACM
  International Conference on Automated Software Engineering}}
  \emph{(\bibinfo{series}{ASE '10})}. \bibinfo{pages}{313–316}.
\newblock
\showISBNx{9781450301169}
\urldef\tempurl%
\url{https://doi.org/10.1145/1858996.1859059}
\showDOI{\tempurl}


\bibitem[Spotbugs(2023)]%
        {spotbugs}
\bibfield{author}{\bibinfo{person}{Spotbugs}.} \bibinfo{year}{2023}\natexlab{}.
\newblock \bibinfo{booktitle}{\emph{{Spotbugs}}}.
\newblock
\urldef\tempurl%
\url{https://github.com/spotbugs/spotbugs}
\showURL{%
\tempurl}
\newblock
\shownote{Accessed: 2023-09-22}.


\bibitem[Tan and Roychoudhury(2015)]%
        {apr-relifix-icse}
\bibfield{author}{\bibinfo{person}{Shin~Hwei Tan} {and} \bibinfo{person}{Abhik
  Roychoudhury}.} \bibinfo{year}{2015}\natexlab{}.
\newblock \showarticletitle{Relifix: Automated Repair of Software Regressions}.
  In \bibinfo{booktitle}{\emph{Proceedings of the 37th International Conference
  on Software Engineering - Volume 1}} \emph{(\bibinfo{series}{ICSE '15})}.
  \bibinfo{pages}{471–482}.
\newblock
\showISBNx{9781479919345}


\bibitem[Touvron et~al\mbox{.}(2023)]%
        {llama-23}
\bibfield{author}{\bibinfo{person}{Hugo Touvron}, \bibinfo{person}{Thibaut
  Lavril}, \bibinfo{person}{Gautier Izacard}, \bibinfo{person}{Xavier
  Martinet}, \bibinfo{person}{Marie-Anne Lachaux}, \bibinfo{person}{Timothée
  Lacroix}, \bibinfo{person}{Baptiste Rozière}, \bibinfo{person}{Naman Goyal},
  \bibinfo{person}{Eric Hambro}, \bibinfo{person}{Faisal Azhar},
  \bibinfo{person}{Aurelien Rodriguez}, \bibinfo{person}{Armand Joulin},
  \bibinfo{person}{Edouard Grave}, {and} \bibinfo{person}{Guillaume Lample}.}
  \bibinfo{year}{2023}\natexlab{}.
\newblock \bibinfo{title}{LLaMA: Open and Efficient Foundation Language
  Models}.
\newblock
\newblock
\showeprint[arxiv]{2302.13971}~[cs.CL]


\bibitem[TreeSitter(2023)]%
        {treesitter}
\bibfield{author}{\bibinfo{person}{TreeSitter}.}
  \bibinfo{year}{2023}\natexlab{}.
\newblock \bibinfo{booktitle}{\emph{{TreeSitter}}}.
\newblock
\urldef\tempurl%
\url{https://tree-sitter.github.io/tree-sitter}
\showURL{%
\tempurl}
\newblock
\shownote{Accessed: 2023-09-22}.


\bibitem[Tufano et~al\mbox{.}(2019)]%
        {apr-tufano-nmt-2019}
\bibfield{author}{\bibinfo{person}{Michele Tufano}, \bibinfo{person}{Cody
  Watson}, \bibinfo{person}{Gabriele Bavota}, \bibinfo{person}{Massimiliano~Di
  Penta}, \bibinfo{person}{Martin White}, {and} \bibinfo{person}{Denys
  Poshyvanyk}.} \bibinfo{year}{2019}\natexlab{}.
\newblock \showarticletitle{An Empirical Study on Learning Bug-Fixing Patches
  in the Wild via Neural Machine Translation}.
\newblock \bibinfo{journal}{\emph{ACM Trans. Softw. Eng. Methodol.}}
  \bibinfo{volume}{28}, \bibinfo{number}{4}, Article \bibinfo{articleno}{19}
  (\bibinfo{date}{sep} \bibinfo{year}{2019}), \bibinfo{numpages}{29}~pages.
\newblock
\showISSN{1049-331X}
\urldef\tempurl%
\url{https://doi.org/10.1145/3340544}
\showDOI{\tempurl}


\bibitem[Tufano et~al\mbox{.}(2021)]%
        {hd-21}
\bibfield{author}{\bibinfo{person}{Rosalia Tufano}, \bibinfo{person}{Luca
  Pascarella}, \bibinfo{person}{Michele Tufano}, \bibinfo{person}{Denys
  Poshyvanyk}, {and} \bibinfo{person}{Gabriele Bavota}.}
  \bibinfo{year}{2021}\natexlab{}.
\newblock \showarticletitle{Towards Automating Code Review Activities}. In
  \bibinfo{booktitle}{\emph{Proceedings of the 43rd International Conference on
  Software Engineering}} \emph{(\bibinfo{series}{ICSE '21})}.
  \bibinfo{pages}{163–174}.
\newblock
\showISBNx{9781450390859}
\urldef\tempurl%
\url{https://doi.org/10.1109/ICSE43902.2021.00027}
\showDOI{\tempurl}


\bibitem[Wang et~al\mbox{.}(2023)]%
        {codet5p-23}
\bibfield{author}{\bibinfo{person}{Yue Wang}, \bibinfo{person}{Hung Le},
  \bibinfo{person}{Akhilesh~Deepak Gotmare}, \bibinfo{person}{Nghi~D.Q. Bui},
  \bibinfo{person}{Junnan Li}, {and} \bibinfo{person}{Steven C.~H. Hoi}.}
  \bibinfo{year}{2023}\natexlab{}.
\newblock \showarticletitle{CodeT5+: Open Code Large Language Models for Code
  Understanding and Generation}.
\newblock \bibinfo{journal}{\emph{arXiv preprint}} (\bibinfo{year}{2023}).
\newblock


\bibitem[Weimer et~al\mbox{.}(2009)]%
        {apr-gp-icse-2009}
\bibfield{author}{\bibinfo{person}{Westley Weimer}, \bibinfo{person}{ThanhVu
  Nguyen}, \bibinfo{person}{Claire Le~Goues}, {and} \bibinfo{person}{Stephanie
  Forrest}.} \bibinfo{year}{2009}\natexlab{}.
\newblock \showarticletitle{Automatically finding patches using genetic
  programming}. In \bibinfo{booktitle}{\emph{2009 IEEE 31st International
  Conference on Software Engineering}}. \bibinfo{pages}{364--374}.
\newblock
\urldef\tempurl%
\url{https://doi.org/10.1109/ICSE.2009.5070536}
\showDOI{\tempurl}


\bibitem[Wong et~al\mbox{.}(2016)]%
        {fl-survey-2016}
\bibfield{author}{\bibinfo{person}{W.~Eric Wong}, \bibinfo{person}{Ruizhi Gao},
  \bibinfo{person}{Yihao Li}, \bibinfo{person}{Rui Abreu}, {and}
  \bibinfo{person}{Franz Wotawa}.} \bibinfo{year}{2016}\natexlab{}.
\newblock \showarticletitle{A Survey on Software Fault Localization}.
\newblock \bibinfo{journal}{\emph{IEEE Transactions on Software Engineering}}
  \bibinfo{volume}{42}, \bibinfo{number}{8} (\bibinfo{year}{2016}),
  \bibinfo{pages}{707--740}.
\newblock
\urldef\tempurl%
\url{https://doi.org/10.1109/TSE.2016.2521368}
\showDOI{\tempurl}


\bibitem[Xuan et~al\mbox{.}(2017)]%
        {apr-nopol}
\bibfield{author}{\bibinfo{person}{Jifeng Xuan}, \bibinfo{person}{Matias
  Martinez}, \bibinfo{person}{Favio DeMarco}, \bibinfo{person}{Maxime Clement},
  \bibinfo{person}{Sebastian~Lamelas Marcote}, \bibinfo{person}{Thomas
  Durieux}, \bibinfo{person}{Daniel Le~Berre}, {and} \bibinfo{person}{Martin
  Monperrus}.} \bibinfo{year}{2017}\natexlab{}.
\newblock \showarticletitle{Nopol: Automatic Repair of Conditional Statement
  Bugs in Java Programs}.
\newblock \bibinfo{journal}{\emph{IEEE Trans. Softw. Eng.}}
  \bibinfo{volume}{43}, \bibinfo{number}{1} (\bibinfo{date}{jan}
  \bibinfo{year}{2017}), \bibinfo{pages}{34–55}.
\newblock
\showISSN{0098-5589}
\urldef\tempurl%
\url{https://doi.org/10.1109/TSE.2016.2560811}
\showDOI{\tempurl}


\bibitem[Yuan et~al\mbox{.}(2022)]%
        {apr-circle-issta-2022}
\bibfield{author}{\bibinfo{person}{Wei Yuan}, \bibinfo{person}{Quanjun Zhang},
  \bibinfo{person}{Tieke He}, \bibinfo{person}{Chunrong Fang},
  \bibinfo{person}{Nguyen Quoc~Viet Hung}, \bibinfo{person}{Xiaodong Hao},
  {and} \bibinfo{person}{Hongzhi Yin}.} \bibinfo{year}{2022}\natexlab{}.
\newblock \showarticletitle{CIRCLE: Continual Repair across Programming
  Languages}. In \bibinfo{booktitle}{\emph{Proceedings of the 31st ACM SIGSOFT
  International Symposium on Software Testing and Analysis}}
  \emph{(\bibinfo{series}{ISSTA 2022})}. \bibinfo{pages}{678–690}.
\newblock
\showISBNx{9781450393799}
\urldef\tempurl%
\url{https://doi.org/10.1145/3533767.3534219}
\showDOI{\tempurl}


\bibitem[Zhang et~al\mbox{.}(2023)]%
        {apr-survey-2023-learning}
\bibfield{author}{\bibinfo{person}{Quanjun Zhang}, \bibinfo{person}{Chunrong
  Fang}, \bibinfo{person}{Yuxiang Ma}, \bibinfo{person}{Weisong Sun}, {and}
  \bibinfo{person}{Zhenyu Chen}.} \bibinfo{year}{2023}\natexlab{}.
\newblock \bibinfo{title}{A Survey of Learning-based Automated Program Repair}.
\newblock
\newblock
\showeprint[arxiv]{2301.03270}~[cs.SE]


\end{thebibliography}

\end{document}